\documentclass[]{interact}

\usepackage{epstopdf}% To incorporate .eps illustrations using PDFLaTeX, etc.
\usepackage[caption=false]{subfig}% Support for small, `sub' figures and tables

\usepackage{natbib}% Citation support using natbib.sty
\bibpunct[, ]{(}{)}{;}{a}{}{,}% Citation support using natbib.sty
\renewcommand\bibfont{\fontsize{10}{12}\selectfont}% Bibliography support using natbib.sty
\usepackage[noperiod]{jabbrv}% Package 'jabbrv' included with template supports abbreviation of journal titles
\usepackage{url}

\theoremstyle{plain}% Theorem-like structures provided by amsthm.sty

\usepackage[ruled,vlined]{algorithm2e}
\SetKw{Break}{break}
\theoremstyle{definition}

\usepackage{hyperref}
\hypersetup{
   colorlinks = true,
   citecolor = blue,
   urlcolor = blue,
   linkcolor = red
}

\theoremstyle{remark}

\usepackage{xcolor}

\newcommand{\atrue}{\delta_j} % active
\newcommand{\arej}{d_j} % rejections
\newcommand{\pch}[1][j]{H_{#1}^{\gamma/s}} % PC null hp
\newcommand{\pcp}[1][j]{p_{#1}^{\gamma/s}} % PC $p$-value
\newcommand{\pcpm}[1][j]{\tilde{p}_{#1}^{\gamma/s}} % PC $p$-value
\newcommand{\rejset}{\mathcal{R}_{\gamma}} % rejection set
\newcommand{\selset}{\mathcal{S}} % rejection set

\begin{document}

\articletype{PREPRINT}% Specify the article type or omit as appropriate

\title{Partial Conjunction Analysis in Neuroimaging: A Comparative Study}

\author{
\name{Monitirtha Dey\textsuperscript{a}, Anna Vesely\textsuperscript{b} and Thorsten Dickhaus\textsuperscript{a,*}\thanks{*Corresponding Author. Email: dickhaus@uni-bremen.de}}
\affil{\textsuperscript{a}Institute for Statistics, University of Bremen, Bremen, Germany; \textsuperscript{b}Department of Statistical Sciences, University of Bologna, Bologna, Italy}
}

\maketitle

\begin{abstract}
Replicability is a cornerstone of science. The partial conjunction (PC) hypothesis testing framework objectively quantifies replicability across disciplines. Although several statistical methodologies for testing PC hypotheses exist, it is not clear which method performs well under which circumstances. In this paper, we consider the PC hypothesis testing problem from a neuroimaging perspective. Identifying the brain regions activated by a specific cognitive task constitutes a central challenge in neuroimaging.
This problem becomes complex when the objective is to evaluate whether activation patterns are consistent across different cognitive tasks or subjects. In this paper, we cast this question as a PC hypothesis testing problem, assessing, for each location in the brain, whether it is activated in at least $\gamma$ subjects, for a pre-specified granularity $\gamma$. In our comparative study, we consider three methods, namely: adaFilter, CoFilter, and a method proposed by Benjamini, Heller, and Yekutieli (BHY). %The simulations elucidate that adaFilter works well for higher values of $\gamma$ whereas the recently proposed CoFilter works well for lower values of it. The BHY method also has comparable power in different settings. 
In equi-correlated simulated data, the BHY procedure tends to outperform the competing methods for high values of $\gamma$, while CoFilter performs well for low values of $\gamma$. In the real-data analysis, CoFilter dominates the other methods for intermediate values of $\gamma$.

\end{abstract}


\section{Introduction}

Replicability is a cornerstone of many scientific disciplines. Replication of a scientific discovery in multiple studies increases the credibility of the finding. Indeed, if a result is consistent across several studies, it is more likely to establish reliable new knowledge. In his famous article, \cite{Ioannidis} argued that most published research findings are neither replicable nor reproducible. \cite{Camerer} evaluated the replicability of 18 laboratory experimental papers in economics (published in the \textit{American Economic Review} and the \textit{Quarterly Journal of Economics} during 2011-2014) and found a significant effect in the same direction as the original study for only 11 among those. According to a 2021 cross-sectional survey of cardiology articles (2014–2019) \citep{Anderson2021}, independent reproduction was practically impossible because 96.6\% of them lacked access to raw data, 98.7\% did not give analysis scripts, and 98.3\% did not include complete study protocols. 

There have also been many calls for changes in statistical methods to alleviate these issues, including suggestions to abandon the usage of statistical significance or to develop formal statistical guidelines for authors. These have led to a dynamic shift in the usage of statistics in various fields. Prestigious publication outlets, e.g., the \textit{Statistical Science} \citep{Carriquiry}, the \textit{Harvard Data Science Review} \citep{Stodden2020}, the \textit{Evolutionary Computation Journal}, and the \textit{ACM Journal of Data and Information Quality} have published special issues on replicability and reproducibility in their respective domains in the last few years. While reproducibility requires detailed study protocols, thorough documentation, and full analytic transparency, replicability can be assessed quantitatively with statistical methods.

\cite{Friston2005} proposed the partial conjunction (PC) hypothesis testing framework to assess replicability quantitatively. Suppose we have $s$ null hypotheses (often referred to as base nulls throughout this work) $H_1, \ldots, H_s$. Given a positive integer $\gamma \leq s$, the PC null hypothesis \citep{benjamini2008screening} states that at most $\gamma-1$ of the $s$ base nulls are false. In the multi-studies setting, every base null represents a test from one study, and rejecting a PC null here implies that the signal is replicated in at least $\gamma$ of $s$ studies. PC hypothesis testing finds applications across scientific domains \citep{Dickhaus2026}. For instance, it can be used to identify brain voxels that are activated either in response to a stimulus in at least $\gamma$ of $s$ individuals, or in response to at least $\gamma$ of $s$ different stimuli \citep{benjamini2008screening}; \cite{SunWei2011} studied the genes differentially expressed in at least $\gamma$ time points, while \cite{Karmakar} identified the outcomes with at least $\gamma$ evidence factors. PC hypothesis testing has also been used to find the features for which the results are replicated in at least $\gamma$ of $s$ studies \citep{bogomolov2023testing, Dickhaus2026}. 

The present work is focused on assessing replicability in neuroimaging. 
Scientifically investigating the topographical organization of the human brain and understanding the unique contribution of its distinct regions has been a major challenge in the neuroscience community \citep{Lazar2002, Lisman2015}. Although many brain regions have been defined, a thorough formalization of each region’s role on human behavior has not been completely elucidated yet \citep{TiCS2018}. Hence, identifying the brain regions activated through a particular cognitive task becomes a problem of pivotal importance in neuroimaging. This challenge is complicated when the interest lies in assessing the consistency of activation patterns across multiple cognitive tasks or subjects, as it requires identifying regions that are reliably engaged across tasks or individuals.

Such investigations often rely on functional magnetic resonance imaging (fMRI) data. Typically, fMRI studies measure brain activation as changes in blood flow, known as the blood oxygenation level dependent (BOLD) signal, under cognitive stimuli. This is usually done at the level of individual voxels, the three-dimensional units that constitute the brain image, by correlating the sequence of stimuli with the BOLD time series in each voxel. A statistically significant (insignificant) correlation then leads to declaring the corresponding voxel active (inactive) \citep{Lindquist2008}.

There exists a plethora of methods for activation localization in neuroimaging. The most prominent and widely used approach is the \textit{cluster-based inference},  which involves identifying brain regions as contiguous supra-threshold clusters and uses random field theory (RFT) to detect activation, assuming that the spatial autocorrelation function has a squared exponential shape. Besides the recent criticisms on the validity of this assumption \citep{Eklund}, the RFT-based cluster inference suffers from low spatial specificity: whenever a cluster is declared active, no information is provided about the precise location or extent of the activation within it \citep{Goeman2018, Woo2014}. To address this issue, \citet{Goeman2018} implemented the simultaneous inference procedure for true discovery proportions (TDPs) introduced by \citet{Goeman2011}, thereby establishing the \textit{all resolutions inference} (ARI) methodology. This approach allows the researcher to apply any data-driven region selection rule and estimate, through lower confidence bounds, the proportion of active voxels within any region - everything from the same dataset. Since the work of \cite{Goeman2018}, several other methodologies have focused on TDP inference in neuroimaging, e.g., \cite{sanssouci, BLAIN, Andreella, Vesely, Preusse2025}. For a discussion of recent proposals, see \cite{Andreella2024}.

In this paper, we address a related but distinct research question: while existing methods for TDP inference typically aggregate information across all subjects into a single $p$-value matrix to quantify regional activation, our focus is on analyzing each subject individually (i.e., their subject-specific $p$-value matrices) to assess the consistency of the signal across subjects. Specifically, for each voxel, our goal is to infer a lower bound on the number of subjects for which that voxel is active. This problem can be conceived as a
%\textit{partial conjunction hypotheses} (PC hypotheses henceforth) \citep{benjamini2008screening}, i.e., we are testing whether a specific brain region is activated in at least $\gamma$ (for some pre-fixed $\gamma$) subjects. 
PC hypothesis testing problem \citep{benjamini2008screening}, where we test whether a specific brain region is activated in at least $\gamma$ (for some pre-fixed $\gamma$) subjects. 
%In this paper, we are interested in a related but different research question: the original ARI aggregates information from all subjects into a single $p$-value matrix, while we study subjects (or equivalently the subject-specific $p$-value matrices) separately.

Towards this, we implement different existing methodologies for PC hypothesis testing and compare their performances in simulation studies and real data analysis. In particular, we examine the adaFilter method of \cite{Owen}, the procedure of \cite{benjamini2009selective} (referred to as BHY henceforth), and the CoFilter procedure of \cite{Dickhaus2026}.

It is mentionworthy that scattered, spread-out signals (in single voxels or very small groups of voxels) are prone to be artifacts \citep{Peitek}. Instead,
topologically contiguous signals forming larger groups of voxels are much more plausible. However, one may still perform voxel-wise inference as a first step, followed by region-wise assessment. \cite{Peitek} combine the original $p$-values 
within atlas-based regions (co-registered with each individual’s anatomical data) to test for the
statistically significant involvement of that region. Thus, testing individual voxels for activation is embeddable into an analysis pipeline which aims at identifying active regions.

This paper is organized as follows. In Section \ref{section2}, we formally introduce the framework and our goal. Section \ref{section3} presents the existing methodologies for testing PC hypotheses. Section \ref{section4} illustrates the empirical performances of the various methods through a simulation study, while Section \ref{section5} contains a data analysis on a real fMRI dataset. We conclude the paper in Section \ref{section6}. Throughout this paper, for any positive integer $n$, $[n]$ denotes the set $\{1, \ldots, n\}$.

\section{Framework \label{section2}}

We think of the brain as a collection of $m$ voxels. Analogously to \cite{Goeman2018}, we assume that for each voxel a test statistic for activation has been computed, which can be converted
into a voxel-wise $p$-value. \cite{Mumford} present how these test statistics or $p$-values are computed in practice. We examine $s$ independent maps (one corresponding to each subject) of $p$-values, each referring to the same $m$ voxels. The $i$'th map, $i\in [s]$, contains the $p$-values
\[(p_{1i},\ldots,p_{ji},\ldots,p_{mi})\]
where $p_{ji}$ corresponds to a certain null hypothesis $H_{ji}$. Here, $H_{ji}$ is the null hypothesis that voxel $j$ is not active in subject $i$ during the task under consideration.

We are interested in studying the replicability of signal detection over the $s$ subjects. The goal is determining, for each voxel, whether there is significant signal in at least a given proportion $\gamma/s$ of subjects, for any granularity $\gamma\in [s]$. As mentioned in the introduction, this inference problem is a PC hypothesis testing problem \citep{Heller2007, benjamini2008screening, bogomolov2023testing, Liang2025powerful, Dickhaus2026}. In the following, we formalize it. 

Consider any voxel $j\in [m]$. The unknown number of subjects in which it is active is given by
\[\atrue = |\{i\in [s]\;:\,H_{ji}\text{ is false}\}|.\]

So, for each $j$, $\delta_{j}$ lies between zero and $s$, both inclusive. For any granularity
$\gamma\in [s]$, one may define the PC  null hypothesis that the voxel is active in less than $\gamma$ subjects:
\[\pch\,:\,\atrue < \gamma\qquad\text{vs}\qquad K_j^{\gamma/s}\,:\,\atrue\geq \gamma.\]
Note that, equivalently, 
\[\atrue = \max\{\gamma\in [s]\;:\,\pch\text{ is false}\}. \]
If $\gamma$ is fixed, we are interested in testing the PC null hypotheses for all voxels $\pch[1],\ldots,\pch[m]$ simultaneously, with some type I error control. The literature on multiple testing has witnessed the development of several notions of type I error rates. However, it was the landmark article by \cite{benjamini1995controlling} which induced a paradigm shift in multiple testing research by proposing the false discovery rate (FDR). Over the last three decades, this has attracted diverse interest from the theoretical point of view, extending and expanding our understanding of simultaneous inference in many directions. The FDR has emerged as the standard type I error criterion in large-scale multiple testing; cf., e.\ g., \cite{Dickhaus}. Since in our applications the multiplicity $m$ is in the thousands, we choose the FDR as the type I error criterion in this work.

In neuroimaging literature, the test with $\gamma=s$ (i.e, whether all the subjects show real effects in voxel $j$) is referred to as the test of \textit{conjunction null} \citep{Friston2005, Nichols2005, Price1997}. It is not clear as to what value of $\gamma$ should be chosen in order to establish consistency across subjects (or replicability, in general) \citep{benjamini2009selective}. However, instead of predefining $\gamma$, we may test in order
the PC hypotheses with $\gamma = 1, 2, \ldots, s$.  Investigating all granularities $\gamma$ simultaneously is the foremost objective of this paper. This leads to obtaining for each voxel $j$ a lower bound for the number of subjects for which the voxel is active:
\begin{align}
\arej=\max\{\gamma\in [s]\,:\,\pch\text{ is rejected}\}. \label{eq:arej}
\end{align}

Analogously to \citet{benjamini2008screening}, we define an overall FDR that we aim at controlling at level $\alpha\in (0,1)$. For each voxel, we have a \textit{discovery} if at least one PC null hypothesis is rejected ($\arej>0$), and we have a \textit{false discovery} if the claim is too strong ($\arej>\atrue$). Then the overall FDR is the expected value of the proportion of overall false discoveries:
\begin{align}
\text{FDR}=\mathbb{E}(\text{FDP}),\qquad\text{FDP}=\frac{|\{j\in [m]\,:\,\arej>\atrue\}|}{|\{j\in [m]\,:\,\arej>0\}|\,\vee\, 1}, \label{eq:FDR}
\end{align}
where the symbol $\mathbb{E}$ refers to expected value under the true, but unknown data-generating process, and $\vee$ denotes the maximum. In literature other measures of error have been used, e.g., the false coverage-statement rate (FCR) \citep{BY2005b, benjamini2009selective}, but in our setting the overall FDR is more intuitive. In particular, the CoFilter method \citep{Dickhaus2026} may test different granularities $\gamma$ with different threshold values $\tau$, as it will be explained in the next sections; in that case, the FCR is not immediately applicable, while the overall FDR can still be computed. 

The existing simultaneous inference literature has several notions of power \citep{Ramsey1978, WTRWH, Dudoit, BretzBook, Dickhaus, Grandhi}. To evaluate the power of the three methods under consideration, we quantify the performance in correctly detecting activation at different granularities. Since replicability requires evidence of an effect in at least two subjects, in most applications the primary interest lies in granularities at least equal to two. Therefore, we define the power at any $\gamma\in\{2,\ldots,s\}$ as the proportion of voxels active in at least $\gamma$ subjects that are correctly identified:
\begin{align}
\beta(\gamma) = \frac{|\{j\in [m]\,:\,\arej\geq\gamma\}|}{|\{j\in [m]\,:\,\atrue\geq\gamma\}|\,\vee\, 1}. \label{eq:power}
\end{align}
This definition captures the expected variation in method performance across different levels of replicability, motivating a $\gamma$-specific measure of power.

\section{Statistical methodologies for PC hypothesis testing \label{section3}}

The main goal is to compute the desired lower bound $\arej$ \eqref{eq:arej} for all voxels with control of the overall FDR \eqref{eq:FDR} at a pre-specified level $\alpha$. We study the performances of three statistical methods: adaFilter \citep{Owen}, the BHY method proposed by \cite{benjamini2009selective}, and the CoFilter method \citep{Dickhaus2026}.

Each of these methods tests the PC hypothesis $\pch$ using the so-called PC $p$-value $\pcp$, for any granularity $\gamma\in [s]$ and any voxel $j\in [m]$. The PC $p$-value is obtained by combining the original $p$-values $p_{j1},\ldots,p_{js}$ from all subjects. More specifically, one may derive $\pcp: [0, 1]^s \to [0, 1]$ in the following way: 
\begin{enumerate}
\item At first, take a global null combination
$p$-value $p_{j}^{1/(s-\gamma+1)} : [0, 1]^{s-\gamma+1} \to [0, 1]$ for $s - \gamma + 1$ hypotheses, that is increasing in each argument.
\item Define 
$$\pcp = \pcp(p_{j1}, \ldots, p_{js})=p_{j}^{1/(s-\gamma+1)}(p_{j(\gamma)},\ldots, p_{j(s)})$$
\end{enumerate}
where $p_{j(1)}\leq\ldots\leq p_{j(s)}$ are the sorted base $p$-values. Therefore, only the $s-\gamma + 1$ largest $p$-values are combined. There exist various $p$-value combining functions \citep{benjamini2008screening, Loughin2004}. This paper considers the Fisher combination function (Section 21.1 in \cite{Fisher1934}):
\begin{align}
\pcp=1-F_{\chi^2_{2(s-\gamma+1)}}\left(-2\sum_{i=\gamma}^s\log p_{j(i)} \right) \label{eq:fisher}
\end{align}
where $F_{\chi^2_{2(s-\gamma+1)}}$ denotes the cumulative distribution function of the chi-square distribution with $2(s-\gamma+1)$ degrees of freedom. The Fisher combination function also has excellent power properties for a wide range of signals \citep{benjamini2008screening, Dickhaus2022}. 

Standard PC hypothesis testing methods relying directly on $p_{j}^{\gamma/s}$ can be overly conservative for testing $H_{j}^{\gamma/s}$. This is because $p_{j}^{\gamma/s}$ can be highly superuniform under the global null, i.e, $\mathbb{P}(p_j^{\gamma/s}\le \nu)\ll \nu$ under the global null \citep{Liang2025powerful}. To overcome this conservativeness, adaFilter and CoFilter, respectively, are leveraging an initial filtering step that retains only PC hypotheses whose PC $p$-values are smaller than a data-driven threshold $\tau$. When $\delta_j = 0$ for almost all $j$'s, the PC $p$-values are conservative and so few fall below $\tau$. Restricting inference to this smaller subset of PC hypotheses substantially reduces the multiplicity of the underlying problem and boosts power \citep{dey2025pch}.

For each granularity $\gamma$, we obtain $m$ PC $p$-values, $\pcp[1],\ldots,\pcp[m]$. All three considered methods employ a multiple testing procedure; here we use the Benjamini–Hochberg (BH) procedure \citep{benjamini1995controlling}. By combining results for different values of $\gamma$, we finally obtain the lower bounds $\arej$ as in \eqref{eq:arej}.

Conceptually, there is an important distinction: BHY guarantees control of the overall FDR \eqref{eq:FDR} across all $\gamma$, whereas the other methods control the FDR only for a fixed, pre-specified $\gamma$. Nevertheless, we examine the robustness of all methods by considering multiple values of $\gamma$ and comparing the resulting $\arej$ for all voxels $j\in [m]$.

\subsection{AdaFilter}

For any granularity $\gamma\in\{2,\ldots,s\}$, \cite{Owen} define the auxiliary (filter) and the PC p-values for all voxels $j$ as
\begin{align}
\label{eq:adafilter1}
a_j^{\gamma/s} = (s-\gamma+1) p_{j(\gamma-1)},\qquad \tilde{p}_j^{\gamma/s}=(s-\gamma+1) p_{j(\gamma)}.
\end{align}

\noindent Then they adopt the following decision rule: $H_{j}^{\gamma / s}$ is rejected if $\tilde{p}_j^{\gamma/s}<t_\alpha$, where
\begin{align}
\label{eq:adafilter2}
t_\alpha=\sup \left\{t \in[0, \alpha]\,:\, \frac{t\cdot |\{j\in [m]\,:\,a_j^{\gamma/s}<t\}|}{|\{j\in [m]\,:\,\tilde{p}_j^{\gamma/s}<t\}| \vee 1} \leq \alpha\right\}
\end{align}
and $|\cdot|$ denotes the size of a set.

Theoretical FDR control is guaranteed for independent $p$-values and at the conservative level 
$\alpha\, c_m$ with $c_m = \sum_{j=1}^{m} 1/m \approx \log m$, while the procedure controls the FDR at level $\alpha$ under weak dependencies asymptotically, as $m \to \infty$. However, \citet{Owen} claim that strong empirical evidence suggests that in practice the adjustment by $c_m$ is generally not needed.

The method is summarized in Algorithm \ref{al:m0} and is implemented in the \texttt{R} package \texttt{adaFilter}.

\begin{algorithm}
\caption{\label{al:m0} Algorithm for adaFilter}
\SetAlgoLined
\KwData{$p$-values $p_{ji}$ for each subject $i\in [s]$ and voxel $j\in [m]$; granularity $\gamma\in \{2,\ldots,s\}$; level $\alpha\in (0,1)$.}
\KwResult{set $\rejset$ of indices of rejected PC null hypotheses $\pch$.}

\For{$j=1,\ldots,m$}{
	compute $a_j^{\gamma/s}$ and $\tilde{p}_j^{\gamma/s}$ as in \eqref{eq:adafilter1}\;
 }
compute $t_\alpha$ as in \eqref{eq:adafilter2}\;
$\rejset=\{j\in [m]\; :\;\tilde{p}_j^{\gamma/s}< t_\alpha\}$\;
\Return $\rejset$\;
\end{algorithm}

\subsection{BHY}
For any $\gamma\in [s]$, the method of \citet{benjamini2009selective} computes global $p$-values $p_j^{1/s}$ as in \eqref{eq:fisher} for all voxels $j$, applies BH, and selects the voxels for which rejections are made. Let $\selset$ be the set of indices of selected voxels. Then, for each $j\in\selset$, the lower bound is computed as the highest $\gamma$ that leads to a rejection of $H_j^{\gamma/s}$:
\[\arej=\max\left\{\gamma\in [s]\; :\; p_j^{u/s}\leq\frac{|\selset|}{m}\alpha\text{ for all } u=0,\ldots,\gamma \right\} \]
if the maximum exists, and $\arej=0$ otherwise (see Algorithm \ref{al:m2}).

\begin{algorithm}
\caption{\label{al:m2} Algorithm for BHY}
\SetAlgoLined
\KwData{$p$-values $p_{ji}$ for each subject $i\in [s]$ and voxel $j\in [m]$; level $\alpha\in (0,1)$.}
\KwResult{lower bounds $d_1,\ldots,d_m$ as in \eqref{eq:arej}.}

$(d_1,\ldots,d_m)=(0,\ldots,0)$

\For{$j=1,\ldots,m$}{
	compute $p_j^{1/s}$ as in \eqref{eq:fisher}\;
 }

apply BH with level $\alpha$ to $\{p_1^{1/s},\ldots,p_m^{1/s}\}$\;
$\selset=\{j\in [m]\;:\; H_j^{1/s}\text{ is rejected}\}$\;
$\theta=|\selset|\alpha/m$\;

\For{$j\in\rejset$}{
	\For{$\gamma=1,\ldots,s$}{
		compute $\pcp$ as in \eqref{eq:fisher}\;
		\lIf{$\pcp\leq\theta$}{$\arej=\gamma$}
		\lElse{\Break}
 	}
 }
\Return $(d_1,\ldots,d_m)$\;
\end{algorithm}

This method controls the overall FDR \eqref{eq:FDR} if the $p$-values of different voxels $p_{1i},\ldots,p_{mi}$ are independent within each map $i$. \citet{bogomolov2023testing} proposed a generalization of this procedure, and proved that it is valid when the following conditions are satisfied: 
\begin{enumerate}
    \item The $s$ $p$-value maps are independent (which is a natural assumption).
    \item The vector of $p$-values for the hypotheses for subject $i$, for every 
 $i$, satisfies the positive regression dependence on a subset (PRDS) property on the subset of true null hypotheses.
 \item The test statistics are continuous.
\end{enumerate}

\subsection{CoFilter}

CoFilter \citep{Dickhaus2026} is a two-stage multiple testing procedure that proceeds as follows: 
\begin{enumerate}
    \item Select the set with PC $p$-values below a selection threshold $\tau$ (either pre-fixed or estimated from the data);
    \item Apply, within the selected set only, a FDR controlling procedure on the conditional PC $p$-values.
\end{enumerate} 

To start with, consider a fixed threshold $\tau\in (0,1)$. For any granularity $\gamma\in [s]$, the method computes the PC $p$-values $\pcp$ as in \eqref{eq:fisher} for all voxels $j$, then selects the voxels for which $\pcp\leq\tau$. Let $\selset_{\tau}$ be the set of selected indices. Subsequently, for all selected voxels $j\in\selset_{\tau}$, it defines the conditional PC $p$-values
\begin{align}
\pcpm = \frac{\pcp}{\tau}
\label{eq:condp}
\end{align}
and applies BH (see Algorithm \ref{al:p1}). The procedure may be repeated for all granularities; for each voxel, $\arej$ is the highest $\gamma$ that leads to a rejection of $H_j^{\gamma/s}$.

\begin{algorithm}
\caption{\label{al:p1} Algorithm for CoFilter with pre-specified threshold}
\SetAlgoLined
\KwData{$p$-values $p_{ji}$ for each subject $i\in [s]$ and voxel $j\in [m]$; granularity $\gamma\in [s]$; level $\alpha\in (0,1)$; thresold $\tau\in (0,1)$.}
\KwResult{set $\rejset$ of indices of rejected PC null hypotheses $\pch$.}

\For{$j=1,\ldots,m$}{
	compute $\pcp$ as in \eqref{eq:fisher}\;
 }
$\selset_{\tau}=\{j\in [m]\;:\; \pcp\leq\tau\}$\;
\For{$j\in\selset_{\tau}$}{
	$\pcpm=\pcp /\tau$ as in \eqref{eq:condp}\;
 }
apply BH with level $\alpha$ to $\{\pcpm,\; :\;j\in\selset\}$\;
$\rejset=\{j\in\selset_{\tau}\; :\;\pch\text{ is rejected}\}$\;
\Return $\rejset$\;
\end{algorithm} 

An alternative version of CoFilter, referred to as the \emph{greedy version}, does not require fixing $\tau$ a priori. For any granularity $\gamma$, it explores different values of $\tau$ and selects the one that leads to the highest number of rejections (see Algorithm \ref{al:p2}).

The conditional PC $p$-values \eqref{eq:condp} are valid if the null $p$-values are uniform and Fisher's combination function is used. \cite{Dickhaus2026} provide conditions guaranteeing that the CoFilter method controls the FDR for any choice of $\tau$ (pre-fixed or greedy). 

\begin{algorithm}
\caption{\label{al:p2} Algorithm for CoFilter with adaptive threshold (greedy version)}
\SetAlgoLined
\KwData{$p$-values $p_{ji}$ for each subject $i\in [s]$ and voxel $j\in [m]$; granularity $\gamma\in [s]$; level $\alpha\in (0,1)$; set of candidate thresholds $\mathcal{T}$.}
\KwResult{set $\rejset^{\text{best}}$ of indices of rejected PC null hypotheses $\pch$.}

$\rejset^{\text{best}}=\emptyset$\;

\For{$\tau\in\mathcal{T}$}{
	$\rejset =$ rejection set obtained from Algorithm \ref{al:p1}\;
	\lIf{$|\rejset|\geq |\rejset^{\text{best}}|$}{$\rejset^{\text{best}} =\rejset$}
 }
\Return $\rejset^{\text{best}}$\;
\end{algorithm}

\section{Simulation study \label{section4}}
In this section, we explore the behavior of the three competing methods through simulations.
In the scenarios considered, the overall FDR appears to be controlled, and we further examine the methods’ performance in terms of power.

In each scenario, we generate independent $p$-value maps for $s$ subjects, each containing $m$ voxels, so that each voxel $j$ is active in a pre-specified $\atrue$ number of subjects. Then we apply the three methods. We simulate data $1{,}000$ times, and for each run we compute the overall FDP \eqref{eq:FDR} and the power $\beta$ \eqref{eq:power}.
%The empirical FDR is computed as the mean of the FDP over the 1000 simulations.
The empirical FDR and empirical power are computed as the mean FDP and mean power, respectively, across the $1{,}000$ runs.

\begin{figure}[!ht]
\centering
\makebox{\includegraphics[width=13cm]{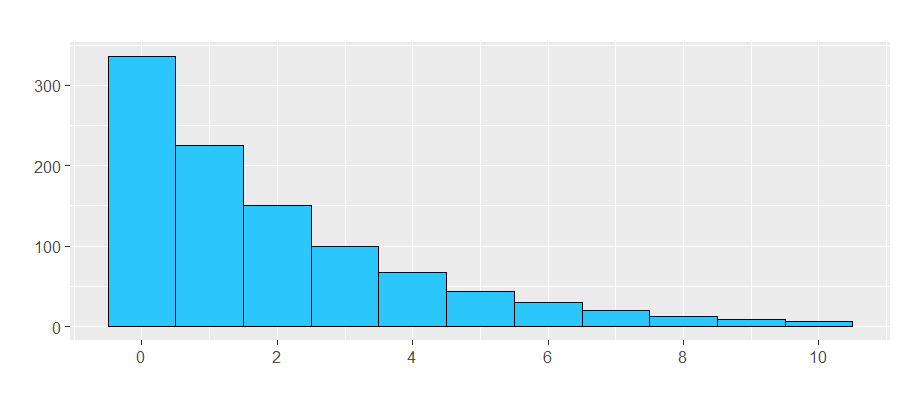}}
\caption{\label{sim1_histdelta} Equi-correlated data: Distribution of $\delta_1,\ldots,\delta_m$, where $\atrue$ represents the number of subjects for which voxel $j$ is active.}
\end{figure}

We set the number of subjects and voxels to $s=10$ and $m=1{,}000$, respectively. The values $\delta_j$ are constructed so that the number of voxels active in exactly $i$ subjects is approximately (up to rounding) proportional to $c^{s-i}$, with $c=1.5$. The resulting distribution is displayed in Figure \ref{sim1_histdelta}.

\cite{Genovese2002neuroimage} remark that strict independence of test statistics across voxels is hard to find in neuroimaging data; however, it is often the case that the correlations tend to be positive. For simplicity, we adopt the equi-correlated setup, the simplest correlation structure with positive correlations. To obtain a single $p$-value map, we first generate $n=50$ independent observations from a multivariate normal distribution (MVN) in dimension $m$: $\boldsymbol{X}\sim\text{MVN}_m(\boldsymbol{\mu},\Sigma_\rho)$, where $\Sigma_\rho$ is an equi-correlation matrix with off-diagonal elements equal to $\rho\in\{0,0.3,0.6,0.9\}$. Some of the entries in $\boldsymbol{\mu}$, selected randomly, are non-zero, with value computed so that the two-sided one-sample t-test with significance level $\alpha$ has a given power $\eta=0.95$.  The $p$-value of voxel $j$ is obtained from a one-sample two-sided t-test for the null hypothesis $H_j:\,\mu_j=0$.

All methods are applied with level $\alpha=0.05$. For CoFilter, we consider the greedy version, which searches over $\tau\in\{0.01,0.02,\ldots,1\}$. 

\begin{table}[!ht]
\caption{\label{tab:sim1_FDP} Equi-correlated data: Empirical overall FDR for different correlation values $\rho$, for adaFilter, BHY, and CoFilter. Target level $\alpha=0.05$. All values in the table are based on $1{,}000$ repetitions.}
\centering
\begin{tabular}{cccc}
\toprule
$\rho$ & adaFilter & BHY & CoFilter\\
\midrule
0 & $0.0351$ & $0.0414$ & $0.0262$\\
$0.3$ & $0.0349$ & $0.0410$ & $0.0256$\\
$0.6$ & $0.0314$ & $0.0391$ & $0.0232$\\
$0.9$ & $0.0222$ & $0.0345$ & $0.0182$\\
\bottomrule
\end{tabular}
\end{table} 

\begin{figure}[!ht]
     \centering
         \includegraphics[width=0.95\textwidth]{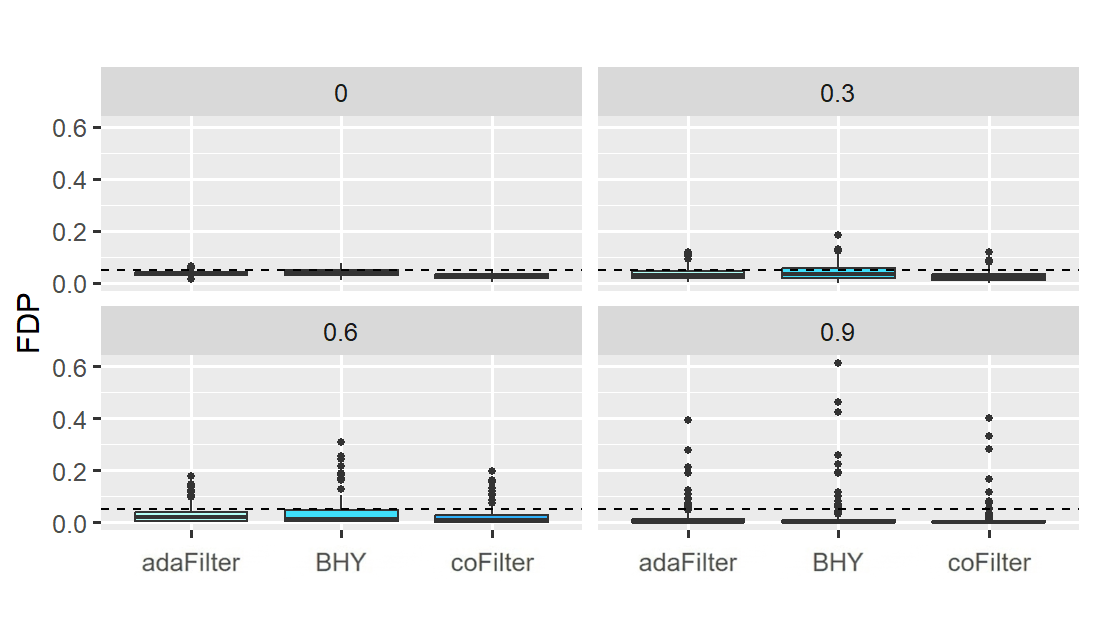}
\caption{\label{sim1_fdp_power} Equicorrelated data: FDP obtained for different correlation values $\rho$, for adaFilter, BHY, and CoFilter. Target level $\alpha=0.05$. All plots are based on $1{,}000$ repetitions. The different considered values of $\rho$ are provided above the subfigures.} \label{sim1_fdp}
\end{figure}

Table \ref{tab:sim1_FDP} presents the empirical overall FDR. The FDR never exceeds $\alpha$; however, the presence of outliers in the right tail of the FDP distribution increases with the correlation parameter $\rho$, reaching values up to $0.6$ for BHY, as illustrated in \autoref{sim1_fdp}.
%In this setting, the adaFilter method tends to be the most powerful.
%Analogously to the FDP, the power is not influenced by the parameter $\rho$ on average, but it becomes more variable as $\rho$ increases.

\begin{figure}[!ht]
\centering
\makebox{\includegraphics[width=13cm]{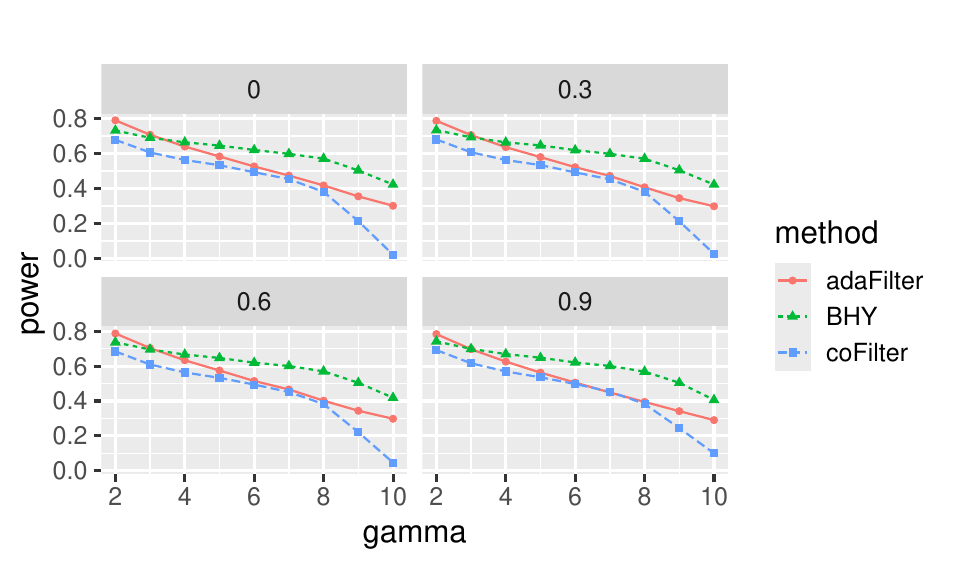}}
\caption{\label{plot_power_sim1} Equi-correlated data: power comparison. All graphs are based on $1{,}000$ repetitions. The considered values of $\rho$ are provided above the subfigures.}
\end{figure}

\autoref{plot_power_sim1} compares the empirical power of the three methodologies. Analogously to the FDR, the average power is not strongly influenced by the correlation parameter $\rho$. Different methodologies achieve the highest power in different ranges of $\gamma$, resulting in several \textit{crossing points}.
%meaning that one may identify gamma-ranges in which each of the procedures is preferable.
%For example, we observe that adaFilter performs much better than CoFilter for large values of $\gamma$, as also mentioned in \cite{Dickhaus2026}.
For example, adaFilter exhibits higher power for $\gamma=2$, while BHY is the most powerful method for larger values of $\gamma$, consistent with the observations of \cite{Dickhaus2026}. Cofilter and adaFilter exhibit similar power for intermediate values of $\gamma$.

\section{Real data analysis}\label{section5}
We compare the three considered methodologies on a real dataset from the Neurovault repository, namely the \emph{Midnight Scan Club data} \citep{GORDONneuron}, available at
%The accession number is
\href{https://identifiers.org/neurovault.collection:2447}{https://identifiers.org/neurovault.collection:2447}. The authors argue that the standard human functional neuroimaging approach needs to be expanded by developing
methods to systematically characterize brain function and organization in single individuals. Individual-specific neuroimaging is critical for determining whether differences in brain organization are behavior-related, disease-dependent, or epiphenomenal. They collected
a large quantity of fMRI data from ten healthy individuals (24-34 years; 5 females). Along with other measurements, BOLD responses to foot, hand, and tongue movements were recorded. In this work, we focus on the contrast between left-hand and right-hand movements.

We downloaded t-statistic maps for the $s=10$ subjects, each with spatial dimensions 
$48\times 64\times 48$. Since a brain mask was not provided, we constructed one by excluding all voxels that had zero values across all subjects. This procedure resulted in $m=82{,}984$ voxels within the mask. For each voxel and each subject, we then computed a two-sided $p$-value by approximating the t-distribution with a standard normal distribution. This approximation was necessary as the degrees of freedom of the t-distribution, mostly dependent on the number of time points in the experiment, were not available.

\begin{figure}[!ht]
     \centering
         \includegraphics[width=0.8\textwidth]{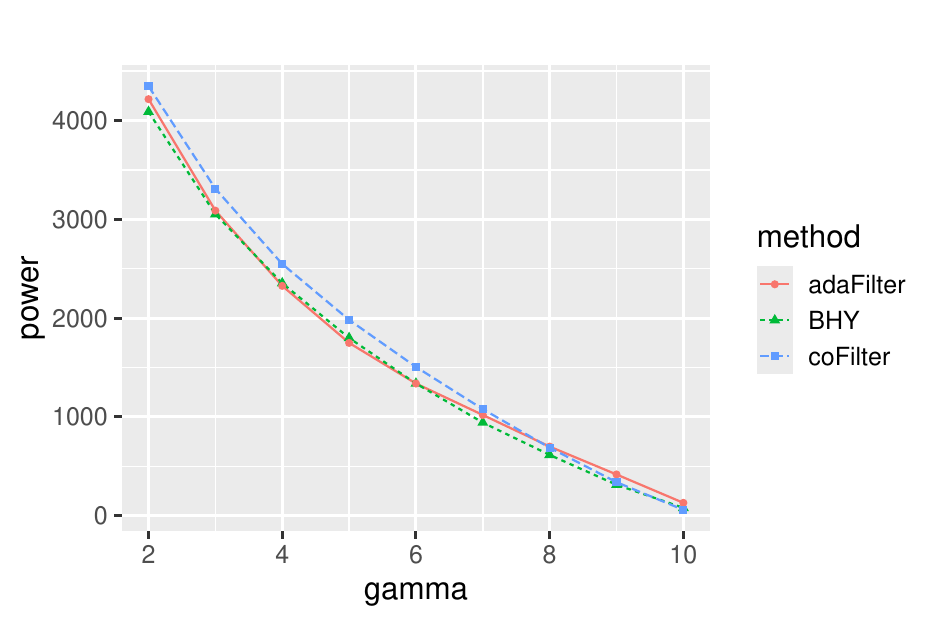}
\caption{\label{brain2} Midnight Scan Club Data: Number of rejections by granularity $\gamma$, where a voxel $j$ is rejected if $d_j\geq\gamma$. %Target level $\alpha=0.05$.
}
\end{figure}

\autoref{tab:brain} and \autoref{brain2} present comparative results of the three methods for this dataset. In particular, for each value of $\gamma$, they display the number of rejections of the corresponding PC null hypotheses, i.e., the number of voxels $j$ with $d_j\geq\gamma$:
\begin{align*}
r(\gamma) = |\{j\in [m]\,:\,\arej\geq\gamma\}|.
\end{align*}
%\autoref{tab:brain} and \autoref{brain2} present comparative results of the three methods for this dataset.
We observe the following. 
\begin{enumerate}
\item BHY is never the most rejective method, for any granularity $\gamma$. 
\item The presence of crossing points in \autoref{brain2} indicates that different procedures perform better over different intervals of $\gamma$.
\item The highest number of rejections is achieved by adaFilter for large replicability levels ($\gamma\in\{8,9,10\}$), and by CoFilter in the other cases.
%Several crossing points are visible in \autoref{brain2}, indicating that the curves of CoFilter and adaFilter intersect multiple times over the range of $\gamma$. \anna{Consequently, distinct intervals of $\gamma$ can be identified in which each procedure performs better.}
%\item For moderate values of $\gamma$, CoFilter is the most rejective method. On the other hand, adaFilter is the least rejective in this region. 
%\item AdaFilter works best for very high values of $\gamma$, e.g., $\gamma=9,10$, similar to the finding in our simulation setting.
%\item \anna{Compared to adaFilter, CoFilter has the largest number of rejections for intermediate values of $\gamma$. This is consistent to the simulation results.}
\end{enumerate}

\begin{table}[!ht]
\caption{\label{tab:brain} Midnight Scan Club Data: Number of rejections by granularity $\gamma$, where a voxel $j$ is rejected if $d_j\geq\gamma$. %Target level $\alpha=0.05$.
%Number of rejected voxels for the three considered methods. Target level $\alpha=0.05$.
For each $\gamma$, the performance of the best method is marked in bold case.}
\centering
\begin{tabular}{cccc}
\toprule
$\gamma$ & adaFilter & BHY & CoFilter\\
\midrule
2 &  4215 & 4087 & \textbf{4346}\\
3 &  3087 & 3051 & \textbf{3309} \\
4 &  2326 & 2353 & \textbf{2547} \\
5 &  1749 & 1799 & \textbf{1982}\\
6 &  1338 & 1338 & \textbf{1506}\\
7 &  1020 & 943 & \textbf{1079}\\
8 &  \textbf{700} & 616 & 689 \\
9 &  \textbf{418} & 314 & 280 \\
10 &  \textbf{131} & 80 & 60\\
\bottomrule
\end{tabular}
\end{table}

{Finally, in \autoref{brain}, representative brain slices are shown for each method, where each voxel $j$ is mapped to the corresponding value of $d_j$. The regions detected by the three methods show substantial overlap across all three dimensions.

\begin{figure}[!ht]
     \centering
     \subfloat[The adaFilter method]{
         \centering
         \includegraphics[width=0.8\textwidth]{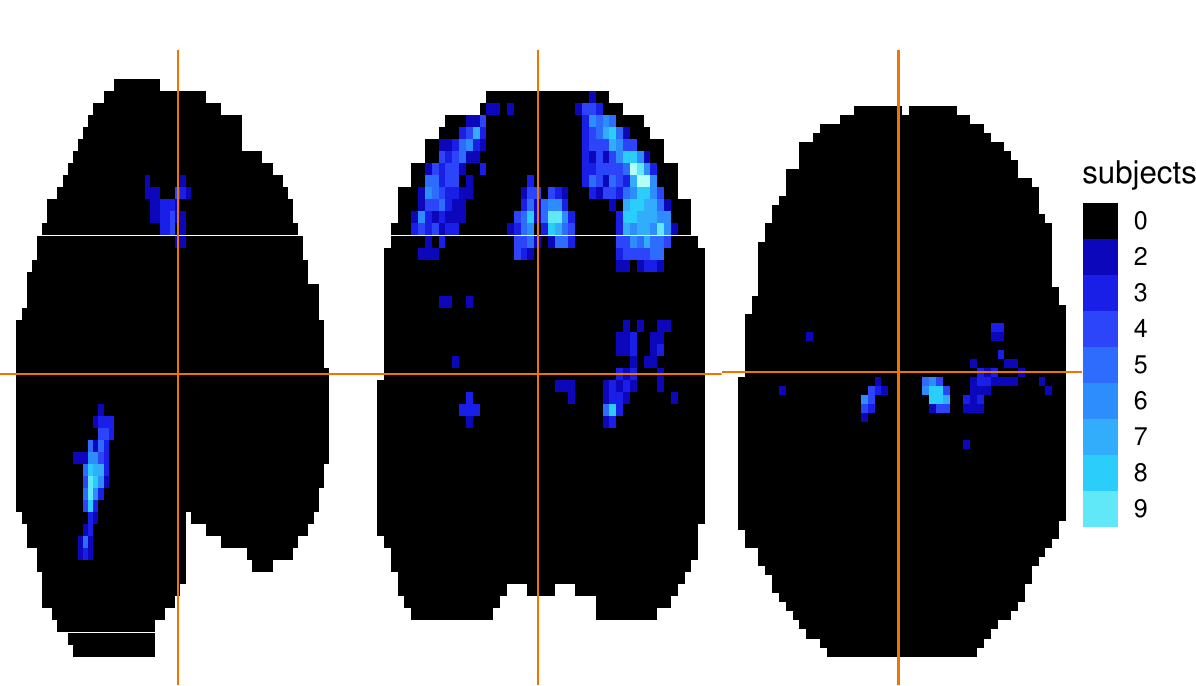}}
     \\
\subfloat[The BHY method]{
         \centering
         \includegraphics[width=0.8\textwidth]{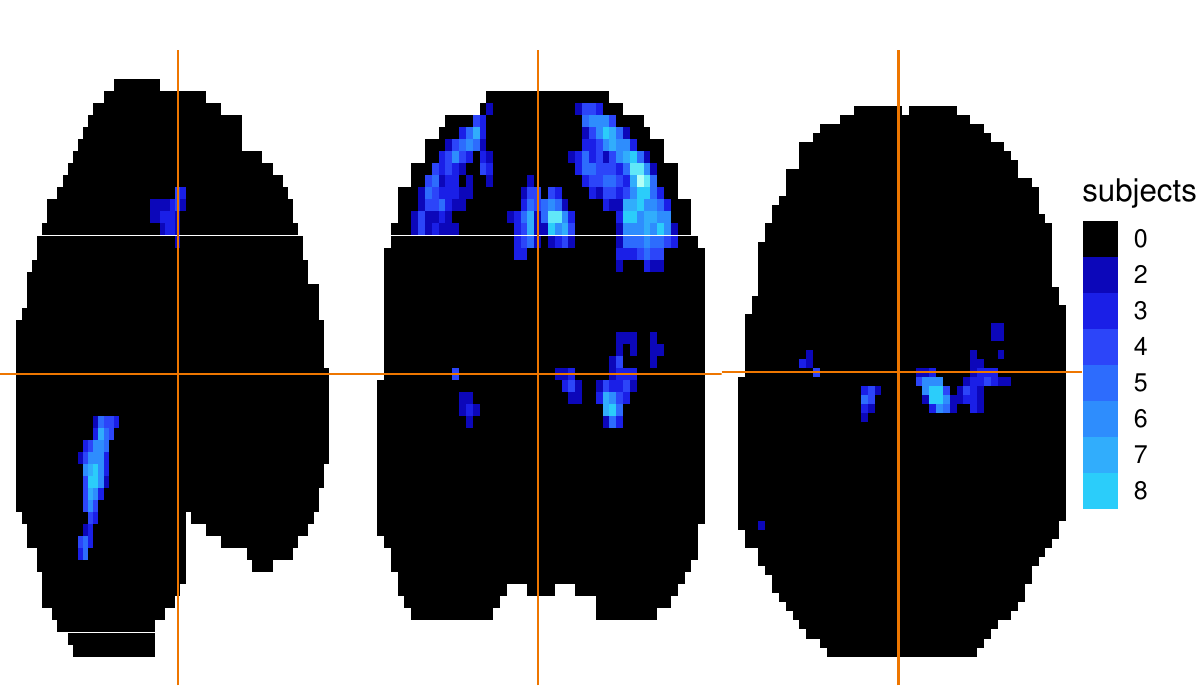}}
     \\
     \subfloat[The CoFilter method]{
     \centering
         \includegraphics[width=0.8\textwidth]{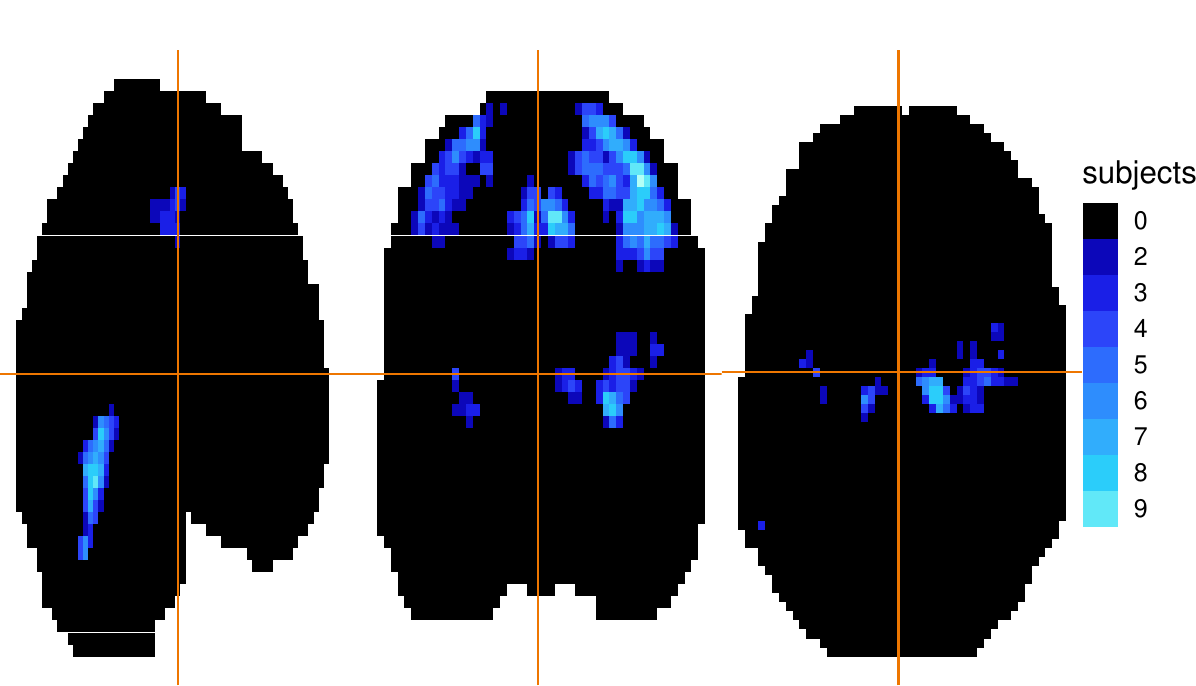}}
\caption{\label{brain} Midnight Scan Club Data: Results $\arej$ for the three methods. Target level $\alpha=0.05$.}
\end{figure}

\section{Concluding remarks \label{section6}} 

Activation localization has been a cornerstone in neuroimaging for decades.  In the present article, we have studied the problem of activation
localization in neuroimaging through the lens of PC hypotheses testing in the sense of identifying
whether a particular voxel is activated (during a specified task) in at least $\gamma$ out of $s$ subjects. Our
foremost technical tools for this identification have been recent advances in the simultaneous
statistical inference literature, namely adaFilter \citep{Owen} and CoFilter \citep{Dickhaus2026}. We have compared these methodologies with the BHY method proposed by \citet{benjamini2009selective} through simulations and by analyzing a real neuroimaging dataset. Simulation results on equi-correlated (synthetic) data ($1{,}000$ voxels) indicated that BHY generally outperforms the competing methods under such a model. In the analysis of the real dataset ($82{,}984$ voxels), CoFilter demonstrated the highest power for most values of $\gamma$, while adaFilter performed best at larger $\gamma$ levels. Therefore, it seems that none of the three methods generally outperforms the others. Instead, it is advisable to choose the data analysis method depending on the model characteristics, especially on the value of the granularity $\gamma$ which is most relevant for the researcher. 
%The simulations elucidate that adaFilter works well for higher values of $\gamma$ whereas the recently proposed CoFilter works well for moderate values of it. The method by \cite{benjamini2009selective} also seems to have comparable power in different settings. 

In our proposed framework, estimation of $\delta_j$ and testing $H_j^{\gamma/s}$ go hand in hand. Several authors have pointed out that in neuroimaging data with a large number of subjects, it may be interesting to estimate $\delta_j$ instead of testing the proportion of non-null hypotheses per voxel $j$ (see, e.g., \cite{benjamini2008screening, Friston1999a, Friston1999b}). Our framework, in a way, achieves that by getting non-trivial lower bounds on the number of activated maps for a particular voxel. 

Although the present paper has focused on applying PC hypothesis testing methodologies in neuroimaging, our framework is quite general and may be used to study similar problems across other disciplines in future work.

\section*{Disclosure statement}

The authors hereby state that they do not have any relevant financial or non-financial competing interest.

\section*{Funding}

The authors gratefully acknowledge financial support by the German Research Foundation (DFG) via Grant No.~DI 1723/5-3. Anna Vesely acknowledges partial financial support from the Italian Complementary National Plan PNC-I.1 “Research initiatives for innovative technologies and pathways in the health and welfare sector” D.D.~931 of 06/06/2022, “DARE - DigitAl lifelong pRevEntion" initiative, code PNC0000002, CUP: B53C22006450001.

% \section*{Notes on contributor(s)}

% An unnumbered section, e.g.\ \verb"\section*{Notes on contributors}", may be included \emph{in the non-anonymous version} if required. A photograph may be added if requested.

% \section*{Nomenclature/Notation}

% An unnumbered section, e.g.\ \verb"\section*{Nomenclature}" (or \verb"\section*{Notation}"), may be included if required, before any Notes or References.

% \section*{Notes}

% An unnumbered `Notes' section may be included before the References (if using the \verb"endnotes" package, use the command \verb"\theendnotes" where the notes are to appear, instead of creating a \verb"\section*").

\bibliographystyle{tfcse}
\bibliography{biblio}

%  The revised 'jabbrv' package files included with this template supports the abbreviation of journal titles throughout the reference section in keeping with ISO/LTWA standards. To use the revised 'jabbrv' package the command line
% \begin{verbatim}
% \usepackage[noperiod]{jabbrv}
% \end{verbatim}
% should be added to the preamble of your .tex file.

% Please include a copy of your .bib file and/or the final generated .bbl file among your source files if your .tex file does not contain a reference list in a \texttt{thebibliography} environment.

%\appendix
%\section{This is the title of the first appendix}
%\section{This is the title of the second appendix}

%\processdelayedfloats %%% See above for an explanation of why this command might be needed here.

%\appendix

\end{document}